\documentclass[preprint,12pt]{elsarticle}

\usepackage{graphicx}%
\usepackage{multirow}%
\usepackage{amsmath,amssymb,amsfonts}%
\usepackage{mathrsfs}%
\usepackage[title]{appendix}%
\usepackage{xcolor}
\usepackage{textcomp}%
\usepackage{manyfoot}%
\usepackage{booktabs}%
\usepackage{algorithm}%
\usepackage{algorithmicx}%
\usepackage{algpseudocode}%
\usepackage{listings}%
\usepackage{longtable}
\usepackage{pdflscape}
\usepackage{url}
\usepackage{adjustbox}
\usepackage{multirow}

\journal{Expert system with applications}

\begin{document}

\begin{frontmatter}



\title{Datasets for Advanced Bankruptcy Prediction: A survey and Taxonomy}


\author[1]{Xinlin Wang}\ead{xinlin.wang@uni.lu} 
\author[1]{Mats Brorsson}\ead{mats.brorsson@uni.lu} 
\author[2]{Zsófia Kräussl}\ead{Zsofia.Kraussl@city.ac.uk}  

\affiliation[1]{organization={Interdisciplinary Centre for Security, Reliability and Trust},
            addressline={29 avenue John F. Kennedy}, 
            city={Kirchberg},
            postcode={1185}, 
            state={},
            country={Luxembourg}}

\affiliation[2]{organization={Faculty of Finance},
            addressline={106 Bunhill Row}, 
            city={London},
            postcode={EC1Y 8TZ}, 
            state={},
            country={United Kingdom}}

\begin{abstract}
Bankruptcy prediction is an important research area that heavily relies on data science. It aims to help investors, managers, and regulators better understand the operational status of corporations and predict potential financial risks in advance. To improve prediction, researchers and practitioners have begun to utilize a variety of different types of data, ranging from traditional financial indicators to unstructured data, to aid in the construction and optimization of bankruptcy forecasting models. Over time, not only instrumentalized data improved, but also instrumentalized methodology for data structuring, cleaning, and analysis. With the aid of advanced analytical techniques that deploy machine learning and deep learning algorithms, bankruptcy assessment became more accurate over time.  However, due to the sensitivity of financial data, the scarcity of valid public datasets remains a key bottleneck for the rapid modeling and evaluation of machine learning algorithms for targeted tasks. This study therefore introduces a taxonomy of datasets for bankruptcy research, and summarizes their characteristics.
This paper also proposes a set of metrics to measure the quality and the informativeness of public datasets.
The taxonomy, coupled with the informativeness measure, thus aims at providing valuable insights to better assist researchers and practitioners in developing potential applications for various aspects of credit assessment and decision making by pointing at appropriate datasets for their studies.
\end{abstract}




\end{frontmatter}



\section{Introduction}\label{sec1}
The prediction of corporate bankruptcy is crucial for the stability of the economic system and the sustainability of business management. In an ever-changing and increasingly competitive global business environment, timely identification of potential financial risks enables companies to implement effective strategic and management measures.

Traditional accounting-based has played a pivotal role in bankruptcy prediction, providing a comprehensive assessment of an enterprise's operational and financial status. 
There are many studies on bankruptcy prediction based on accounting-based data, and most of the studies focus on the model itself to explore how to use a better model to improve the prediction based on current data\cite{altman1994corporate,wu2010comparison,tsai2014comparative,mihalovic2016performance}.

However, publicly available accounting-based data is limited due to several reasons.  First, the sensitivity of financial data often restricts full disclosure. Second, as only listed companies are obliged to disclose financial reports, it is difficult to obtain relevant data on small and medium-sized enterprises. Last, but not least, the lack of uniformity in financial report formats complicates the collection, cleaning, and structuring of large datasets, increasing the likelihood of noisy and redundant data.
In response to these limitations, the research and professional communities are increasingly turning to machine learning and deep learning techniques, referred to in this paper as advanced bankruptcy prediction methods.  We consider these methods as instruments to enhance and increase the quality and \textit{informativeness} of the dataset for a better performance for bankruptcy prediction.

Some studies have integrated macroeconomic data with accounting-based variables to capture some signals that cannot be detected by traditional indicators, enabling earlier warning of bankruptcy risk and providing more time and space for corporate decision-making\cite{altman1968financial,lasheras_2012,Svabova_Michalkova_Durica_Nica_2020,Wang_Ma_Yang_2014,Chuang_2013,Kim_Byeong_Mun_Bae_Byeong_Mun_2018,Alam_Shaukat_Mushtaq_Ali_Khushi_Luo_Wahab_2021,Casado_2013}.Other research has identified new predictors in emerging areas, such as textual data and relational data, bringing new understanding and research directions to both industry and academia~\cite{kou_xu_peng_shen_chen_chang_kou_2021,Jo_Shin_2016,zheng_lee_wu_pan_2021,Lukason_Andresson_2019}. By combining data from various fields—including financial, market, and social aspects—a more comprehensive understanding of the business ecosystem can be achieved, enriching the context for bankruptcy prediction. Given the multifaceted influences on business operations, utilizing multiple datasets helps address this complexity and build more robust predictive models.

Currently, there have been many reviews on bankruptcy prediction modeling (see~\cite{shi2019overview, kumar2007bankruptcy,kovacova2019systematic,jones2023literature,clement2020machine,appiah2015predicting,alaka2018systematic, bellovary2007review}), and they summarize the modeling methods applied to bankruptcy prediction from various perspectives, ranging from traditional techniques to deep learning model applications. However, as the authors of this paper argue, the effectiveness of the model is highly dependent on the quality of the input data. The results of the comparison of the model effectiveness may become obsolete due to the quality differences of the input data. This paper therefore aims at providing a taxonomy by categorizing the datasets that can be found and instrumentalized by scientific publications. By offering the metrics to evaluate the effectiveness of various datasets for bankruptcy prediction, we are able to better understand and collect data that can help predict bankruptcy, so that we can understand the signs and trends and take effective measures ahead of time, thus reducing the risk of bankruptcy.

Compared to other reviews, this paper is dedicated to a comprehensive and in-depth study of multiple datasets applied to corporate bankruptcy prediction, providing a more holistic perspective on the integration of datasets across disciplines. Our contribution has three important pillars, as follows:

\begin{itemize}
    \item To the best our knowledge, this paper is the first attempt to survey topical research papers and to define the taxonomy of bankruptcy prediction datasets that are instrumentalized for research.
    \item The developed taxonomy provides criteria for classifying and comparing different datasets, allowing researchers to obtain a more focused view of relevant datasets, and to cater their research needs accordingly.
    \item The proposed set of metrics evaluates the quality and the informativeness of the datasets, supporting therefore researchers in collecting the most appropriate datasets for bankruptcy prediction.  
\end{itemize}

The remainder of the paper is organized as follows. First, in Section ~\ref{sec2} we describe the manually collected data for our research and we introduce our taxonomy. Section ~\ref{sec3} narrows down our data sample universe to the publicly available datasets, and introduces our measure that we develop to address the informativeness of datasets. Section ~\ref{sec4} concludes with a reflection on our findings.

\section{Data and first results: Survey and taxonomy}\label{sec2}
The first bankruptcy prediction model appeared in 1968 in the seminal work of Altman~\cite{altman1968financial}. He developed the famous Z-Score model to predict bankruptcy based on the following five financial ratios: working capital/total assets, retained earnings/total assets, earnings before interest and taxes/total assets, market value equity/book value of total debt, and sales/total assets. The model was shown to be successful in providing bankruptcy predictions and served as a foundation for subsequent researchers to further refine bankruptcy models. Numerous scholars and researchers have continued to refine and develop various types of bankruptcy prediction model ever since, narrowing down the general focus across different industries and countries~\cite{chouhan2014predicting,anjum2012business,hauschild2013altman,cindik2021revision,bellovary2007review,kovacova2019systematic}. With the emerge of advanced analytical techniques that thrived on machine learning and vast data availability, more diverse set of data variables were applied to predict bankruptcy. Has bankruptcy research become more efficient as a result of data diversity? And what type of data has become available and instrumentalized for bankruptcy research? This latter question underpins the motivation for this paper.  

To arrive at a sound taxonomy, we first assess and collect datasets comprehensively covering the scientific domain of bankruptcy prediction. To ensure an unbiased environment for data collection, we used Google Scholar\footnote{https://scholar.google.com/}. We executed our search using "bankruptcy prediction" and "insolvency prediction" as the primary query terms to identify and collect the relevant datasets, motivated by the seminal work of Bellovary et al.~\cite{bellovary2007review} that discusses an in-depth survey on different methods for bankruptcy prediction.
Notably, we excluded the term "financial distress" to avoid conflating it with bankruptcy, as the former pertains to operational challenges, whereas the latter involves deliberate asset protection decisions by companies\cite{platt2006understanding}.
The study~\cite{puro2019financial} also supports this view that the financial ratios used in the financial distress prediction model don't contribute to bankruptcy prediction.

We identified relevant papers using Google Scholar based on the following criteria: (1) relevance to bankruptcy or insolvency prediction, (2) detailed dataset descriptions, (3) the adoption of machine learning or deep learning methods for model training, and, and (4) publication within the past decade (2013-2023) or representative datasets before 2013. This ten-year timeframe was selected to focus on recent developments in big data and machine learning and their implications for bankruptcy research. 

In consideration of maximizing the retention of diverse datasets, to further restrict our investigation, we considered only important works as assessed by the (1) number of citations and (2) the quality of publication venue, i.e. published in ranked conferences and journals. To maintain the diversity of the datasets that is selected for bankruptcy research, thus to create a rather robust environment for our assessment, we also included papers that were operating unique, problem-specific datasets yet achieved less citations.   Ultimately, we searched for and included all papers that employed machine learning and deep learning techniques for bankruptcy prediction, independent of the type of dataset that were instrumentalized for the analysis. Our survey resulted therefore in a diverse selection of 47 papers in total, including both public and private, i.e. proprietary, datasets that combined both quantitative (i.e. numeric) and qualitative (i.e. text) variables. To assess and to analyze the used data, we further needed to restrict our analysis on research papers that instrumentalized publicly available, i.e. published datasets. We summarize the findings of our survey in the following subsections.  

\subsection{Survey of public available datasets used by advanced bankruptcy prediction}\label{subsec2.1}

After manually reviewing the papers found, we collected information to describe the characteristics of the instrumentalized datasets for bankruptcy prediction in as much detail as possible. Table~\ref{summary} structures the datasets of the collected 47 papers along the following variables: 
(1) publication year, (2) number of samples that the study uses, (3) bankruptcy rate, (4) number of features that is applied in the model, (5) data type, (6) data source, (7) publicly available? (8) fee applied?. 
We use the slash symbol "/" to indicate where information is not specifically retrievable from a particular paper. Even though that many studies claim the datasets that they use were publicly available, some datasets are only conditionally accessible, such as with subscription fee. The authors of this paper note that it may actually cause a barrier for reviewers to revisit a particular study and/or to use a particular dataset. In the followings we introduce and describe the datasets that are instrumentalized in research, and are claimed to be publicly available with free access.

\subsubsection{American dataset}\label{subsubsec2.1.1}
This dataset was published by Lombardo et al.\footnote{\url{https://github.com/sowide/bankruptcy_dataset}}~\cite{lombardo2022machine} in 2022. The authors collected accounting data from 8262 different companies in the period between 1999 and 2018 related to the public companies in the US stock market. A company is labeled "Bankruptcy" (1) if it filed for bankruptcy under Chapter 11 or Chapter 7 of the Bankruptcy Code in the subsequent fiscal year; otherwise, it is labeled "Alive" (0).  The dataset comprises 78,682 firm-year observations without missing or synthetic values, divided into training (1999-2011), validation (2012-2014), and test (2015-2018) sets.. 

\subsubsection{Polish dataset}\label{subsubsec2.1.2}
This dataset is published in the UCI machine learning repository, which can be accessed and downloaded\footnote{\url{https://archive.ics.uci.edu/dataset/365/polish+companies+bankruptcy+data}}~\cite{misc_polish_companies_bankruptcy_data_365}. The dataset focuses on the bankruptcy prediction of Polish companies, collected from the Emerging Markets Information Service (EMIS\footnote{\url{http://www.securities.com}}) covering bankruptcies from 2000-2012 and active companies from 2007-2013. This dataset doesn't show the original value of the financial statements but contains 64 financial ratios as model training features calculated from the financial statements. The response variable Y uses 0 to represent the company does not bankrupted and uses 1 to represent the company go bankruptcy.
Five classification cases are based on the forecasting period. The nth year means the data contains financial rates from the nth year of the forecasting period and the corresponding class label that indicates bankruptcy status after (5-n) years. 

\subsubsection{Taiwanese dataset}\label{subsubsec2.1.3} 
This dataset is published in the UCI machine learning repository, and it can also be found in Kaggle\footnote{\url{https://archive.ics.uci.edu/dataset/572/taiwanese+bankruptcy+prediction}}~\cite{misc_taiwanese_bankruptcy_prediction_572}. The dataset, spanning from 1999 to 2009, was collected from the Taiwan Economic Journal. Company bankruptcy is defined according to the business regulations of the Taiwan Stock Exchange. The dataset comprises 6,819 instances and 95 variables, with the response variable being binary: 220 companies are labeled as bankrupt and 6,599 as non-bankrupt, resulting in a bankruptcy rate of approximately 3.23\%.
A notable advantage of this dataset is the absence of missing values and the high quality of the variables, which comprehensively cover important financial ratios. However, a significant limitation is its relatively small size, which may be insufficient for training machine learning or deep learning models effectively.

\subsubsection{SMEsD}\label{subsubsec2.1.4}
This dataset was published by Wei et al. in 2024 on Github\footnote{\url{https://github.com/shaopengw/comrisk}}~\cite{zhao_wei_guo_yang_chen_li_zhuang_liu_kou_2022}. This dataset specially contains the information of lawsuit, which is very rare public available data. SMEsD covers 3,976 SMEs and affiliated individuals in China, covering the period from 2014 to 2021. This database forms a comprehensive enterprise knowledge graph linking all enterprises and their related individuals by their basic information and lawsuit records from 2000 to 2021. Basic information for each company includes registered capital, paid-in capital, and establishment date. Lawsuit records provide details such as the involved plaintiff, defendant, subjects, court level, outcome, and timestamp. It can be used to split the out of time dataset for testing.

\subsubsection{HAT}\label{subsubsec2.1.5}
This dataset was collected in early October 2020 and published in 2021~\cite{zheng_lee_wu_pan_2021}, also available on Github\footnote{\url{https://github.com/hetergraphforbankruptcypredict/HAT}}. The authors curated a real-world dataset encompassing the board member network and the shareholder network of 13,489 companies in China. Data was sourced from various public platforms. Specifically, they randomly sampled 1,000 companies located in the same city that experienced bankruptcy in 2018. They then expanded the network by capturing information on all shareholders and board members associated with these firms. This process was repeated twice, resulting in the augmentation of the original 1,000-node network to a more extensive network comprising 13,489 nodes.

\subsubsection{Other publicly available datasets with a subscription fee}\label{subsubsec2.1.6}
\paragraph{Bureau Van Dijk}
Bureau Van Dijk, a Moody's Analytics company, provides a range of business intelligence and financial datasets that are widely used in various industries for decision-making, risk assessment, and research. Some notable datasets offered by Bureau van Dijk include:
\begin{itemize}
    \item Orbis: A comprehensive global company database featuring information on private and public companies worldwide, including detailed financial statements, ownership structures, and key business details.
    \item Amadeus: Focuses on European companies, this datasets provides detailed financial information and company profiles, which is useful for financial analysis, market research, and assessing business relationships.
    \item Fame: A UK and Irish-based database offering financial information on companies operating in the United Kingdom and Ireland, including financial statements, directors' details, and ownership structures.
    \item Zephyr: A global mergers and acquisitions (M\&A) database, offering information on deals, rumors, and market rumors, valuable for analyzing M\&A trends, deal structures, and market dynamics.
    \item Mint Global: Offers company information and financials for businesses globally, combining data from various sources. Useful for assessing business risk, conducting due diligence, and market research.
    \item TP Catalyst: Specializes in transfer pricing documentation and compliance data. Helps businesses navigate international transfer pricing regulations.
    \item Osiris: Focuses on emerging markets, providing financial information on companies in Asia, Latin America, the Middle East, and Africa. Supports financial analysis and risk assessment in these regions.
    \item Belfirst: A Belgium and Luxembourg-based database offering financial information on companies operating in Belgium and Luxembourg, and its datasource is National Bank of Belgium and Creditreform Luxembourg.
\end{itemize}

In addition, Belfirst and Fame datasets include relational data for bankruptcy prediction, as introduced by E.Tobback et al.~\cite{tobback2017bankruptcy}. They collected data from 2011 to 2014 covering more than 400,000 Belgian SMEs and 2,000,000 UK SMEs. This relational data, based on shared directors/managers, creates a directed graph where nodes represent companies and edges represent links between them.

These datasets from Bureau van Dijk are utilized by financial institutions, corporations, researchers, and analysts to gain insights into company performance, conduct market research, and manage risks associated with business operations and investments. 

\paragraph{Compustat} 
Compustat datasets are comprehensive financial databases that provide detailed information on publicly traded companies. These datasets are widely used by researchers, analysts, and financial professionals to conduct financial analysis, modeling, and research. Compustat is a product of Standard \& Poor's (S\&P) and is considered one of the leading sources of financial data.

Key features and components of Compustat datasets include:

\textit{Financial Statements}: Compustat provides detailed financial statements, including income statements, balance sheets, and cash flow statements. These statements offer a comprehensive view of a company's financial performance over time.

\textit{Ratios and Metrics}: The datasets include various financial ratios and metrics that help in evaluating a company's liquidity, profitability, and overall financial health. Common ratios such as return on equity (ROE), debt-to-equity ratio, and earnings per share (EPS) are included.

\textit{Segment Data}: Compustat datasets often include segment-level information, allowing users to analyze the performance of different business segments within a company.

\textit{Stock Market Data}: Information related to stock prices, trading volumes, and market capitalization is available, enabling the analysis of a company's stock performance and market trends.

\textit{Ownership Data}: Compustat provides data on institutional ownership, allowing users to understand the distribution of a company's shares among different institutional investors.

\textit{Corporate Governance}: Some Compustat datasets include information on corporate governance, executive compensation, and board composition, providing insights into the management structure of companies.

\textit{Global Coverage}: While the primary focus is on U.S. companies, Compustat also includes data on international companies, expanding its coverage to a global scale.

Researchers and analysts use Compustat datasets to perform financial modeling, conduct valuation analyses, and gain insights into industry trends. The data is valuable for academic research, investment analysis, and strategic decision-making within the business and financial sectors.

\paragraph{Thomson Reuters Datastream database}
Thomson Reuters Datastream is a global financial data platform developed and delivered by Thomson Reuters, one of the world's leading information service providers. The platform is focused on providing a wide range of economic, financial, and market data to professional investors, financial professionals and researchers. This dataset can be publicly accessed with a subscription fee. The key indicators of macroeconomic from Thomson Reuters Datastream are \textit{Gross Domestic Product}, \textit{Inflation Rate}, \textit{Interest Rates}, \textit{Trade Balance}, \textit{Money Supply (M1, M2, M3)}, \textit{Consumer Confidence Index} and so on.

\subsection{Taxonomy of datasets used for advanced bankruptcy prediction}\label{subsec2.2}
It is a well-known fact in data science that the quality of the data determines the upper boundary of the model performance. Still, despite the importance of recognizing and acknowledging quality measures of the used data, as we argue, bankruptcy prediction research using machine learning has been dominated by a strong, model-driven focus. Numerous scholarly papers have been reviewing and addressing the performance of bankruptcy prediction models~\cite{appiah2015predicting,alaka2018systematic,shi2019overview,clement2020machine,jones2023literature}. Both Alaka et al.~\cite{alaka2018systematic} and Clement et al.~\cite{clement2020machine} found that no single tool or model consistently outperforms others, suggesting that a hybrid approach may be more effective. The other paper points out that advanced machine learning methods appear to have the greatest promise for future research on firm failure~\cite{jones2023literature}. In contrast to these scientific works, we analyze the various datasets that are instumentalized for bankruptcy prediction. To our knowledge, this is the first study in the data science literature devoted to this subject. 

After manually reviewing the papers we detected a feature-driven pattern of the applied datasets, allowing us to classify the research papers on bankruptcy prediction. We conceptualized and summarized our findings by a \textit{taxonomy}. Based on the keywords and the title of the surveyed papers, combined with the leading variables of the applied methodology for bankruptcy prediction, we classified the employed datasets into the following five categories: (1) accounting-based, (2) market-based, (3) macroeconomic, (4) relational, and (5) non-financial. Table~\ref{taxonomy} structures and defines the different categories of our taxonomy. Below we summarize and define these categories in detail. Additionally, Table~\ref{summary} in the Appendix summarizes the reviewed papers and describes their important data-driven features for bankruptcy prediction. 

\begin{landscape}
\begin{table}[htbp] 
  \centering
  \caption{A taxonomy of the datasets according to the type of data}\label{taxonomy}
  \footnotesize
\begin{tabular}{p{3cm}p{5cm}p{5cm}p{5cm}} 

    \hline
    Taxonomy & Description & Key Indicators & Usage \\
    \hline
    Accounting-based indicators & Financial statements, including income statements, balance sheets, and cash flow statements, provide a detailed account of a company's financial performance and position. & Ratios such as liquidity ratios (e.g., current ratio), profitability ratios (e.g., return on equity), and leverage ratios (e.g., debt-to-equity ratio) are commonly derived from financial statements. & Analysts use these indicators to assess a company's ability to meet short-term and long-term obligations, profitability, and financial stability. \\
    Market-based indicators & Market-based indicators reflect the sentiment of investors and the overall market perception of a company's value. & Stock prices, credit ratings, and bond yields are examples of market-based indicators. Changes in these indicators may signal financial distress. & Analysts incorporate market data to capture external perceptions and reactions to a company's financial health, which may not be fully reflected in financial statements alone. \\
    Macroeconomic factors & Macroeconomic factors encompass broader economic conditions that can impact the financial stability of companies across industries. & Economic indicators such as GDP growth, inflation rates, and interest rates are considered. These factors can influence a company's performance irrespective of its internal financial management. & Understanding the macroeconomic environment helps analysts assess the external forces that may contribute to financial distress. \\
    Relational data & Relational data pertains to the interconnectedness and collaborations that a company has with external entities. This includes partnerships, customer relationships, and supply chain dependencies. & Metrics such as the strength of business partnerships, customer satisfaction levels, and the stability of the supply chain can serve as crucial indicators in evaluating a company's relational health. Additionally, the degree of interconnectedness within an industry or market can be a relevant consideration. & In the context of bankruptcy prediction, understanding relational data provides insights into the company's resilience in the face of external shocks. Examining the strength and stability of relationships with partners and customers helps assess the company's ability to weather economic downturns or industry challenges.  \\
    Non-financial data & Non-financial data includes qualitative information about a company, such as management quality, industry trends, and competitive positioning. & Corporate governance practices, technological advancements, and regulatory changes can be important considerations in assessing a company's resilience. & Non-financial data provides a holistic view of a company's operations and strategic positioning, contributing to a more comprehensive bankruptcy risk assessment. \\
    \hline
  \end{tabular}
\end{table}
\end{landscape}


\subsubsection{Accounting-based datasets}\label{subsec2.2.1}
Traditional financial data is one of the most commonly used data sources in bankruptcy forecasting. These data include financial statements such as income statements, balance sheets, and cash flow statements, as well as data related to financial metrics such as operating income, net profit, and return on assets. These data sources provide important clues about the financial health of a company and are often used to build bankruptcy prediction models based on ratio analysis and statistical modeling.

\subsubsection{Market-based datasets}\label{subsec2.2.2}
The surveyed papers that use market-based data instrumentalize proprietary datasets, provided by \textit{Taiwan Stock Exchange Corporation}, \textit{Bloomberg}, \textit{Center for Research in Security Prices (CRSP) dataset}, \textit{Johannesburg Stock Exchange}, \textit{Korea Stock Exchange KOSPI Index} and \textit{Federal Deposit Insurance Corporation}. The applied methods of these papers, however, provide an indirect signal on the used, market-based indicators for bankruptcy prediction.Some of the recurring features of market-based datasets are stocck price performance, market capitalization, trading volume, market-to-book ratio, credit default swap, volatility of stock returns, liquidity ratios, dividend yield, and market sensitivity.

\subsubsection{Macroeconomic factors}\label{subsec2.2.3}
Macroeconomic data sources include data related to the industry and macroeconomic environment, such as industry growth rates, gross domestic product (GDP), interest rates, and inflation rates. These data can help bankruptcy forecasting models to better consider the impact of the external environment on the business, thus improving the accuracy of the forecast.

\subsubsection{Relational data}\label{subsec2.2.4}
Enterprises are an important part of society and have a network of relationships in society that encompasses a wide range of relationships, including with other enterprises, government agencies, non-profit organizations, customers, suppliers, employees and communities. These relationships have a significant impact on the success and sustainability of the enterprise, and they also reflect the state of the enterprise. We also find the transactional data is an important part to form relation between companies\cite{kou2021bankruptcy,le2018oversampling}, however, unfortunately there is no publicly available data upon that.

\subsubsection{Non-financial data}\label{subsec2.2.5}
We define this category to add all other types of, typically descriptive, data, as in ~\cite{altman2015financial}. Based on the surveyed papers, the majority of instrumentalized datasets are private data sources, however, more studies instrumentalize annual reports of publicly listed companies. As the consequence of the further development of different language models, more studies started paying attention to the annexes of such annual reports. The information that is extracted from these annexes can serve as sentiment variables for bankruptcy prediction, as in ~\cite{mai2019deep}.

By reviewing the features of the applied methodologies of the papers, we can indirectly refer to the characteristics of non-financial datasets. According to our review, fundamental information about a company, such as registration date, location, legal form, ownership structure, product information, or the service portfolio, is the most frequently used non-financial data for bankruptcy prediction. Furthermore, company governance indicators, such as company's management ability, business feasibility, technical ability, are as well crucial elements for bankruptcy assessment. As the reviewed papers~\cite{Ciampi_2015, liang2016financial} argue, governance indicators can reflect the business status and signalize future development of a company. There are also some studies that use tax-based variables or legal (i.e. lawsuit-related) data to predict bankruptcy. Due to data sensitivity, such data is typically owned by private entities.

\begin{figure}[htbp]
  \centering
  \includegraphics[width=\linewidth]{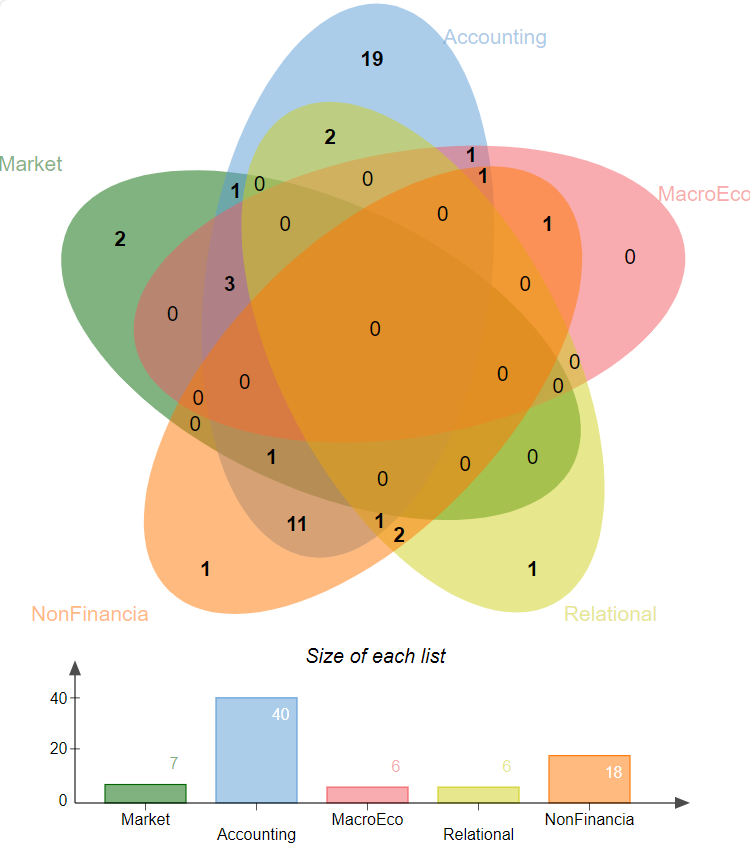}
  \caption{Distribution of survey papers according to taxonomy}
  \label{venn}
\end{figure}

Applying our taxonomy on the total set of surveyed papers, Figure~\ref{venn} depicts the distribution of different dataset types. As our findings show, accounting-based data is still the most commonly utilized data source, which accounts for the vast majority of the reviewed papers. From the in total 47 reviewed studies, 40 papers used accounting-based data. What indicates, however, the emerging trend of using mixed datasets for bankruptcy prediction is that 19 papers out of 40 instrumentalized accounting-based data as the only resource. The remaining 21 studies used other types of datasets to supplement the observations to predict bankruptcy. Only two studies relied purely on market-based data for modelling, and five papers combined market-based data with other data types. From this Venn graph, we know that most papers are rely on a single data type to do the research. It's may because of the difficulty to do the data fusion.

With the development of graph neural networks, relational data has been gradually applied to bankruptcy prediction models in recent years. Non-financial data has become a very broad category, including not only company information on board and management structure and on annual reports, but also on regulatory and compliance data, such as taxation and litigation, as discussed earlier. Our findings indirectly confirm that research methodologies for bankruptcy prediction have been indeed adopting to the available data variables and applied them accordingly. In total we found 24 papers that use either relational (6 papers) or non-financial (18 papers) data. Most frequently, these datasets were applied in combination with accounting data. There are four papers that exclusively use these datasets for bankruptcy prediction.

\section{Experiment and impact: Data quality and informativeness for bankruptcy prediction}\label{sec3}
The presented survey summarizes different datasets that are deployed for advanced bankruptcy prediction, resulting in a taxonomy to structure and to address their main characteristics. As we further argue, this taxonomy enables setting up an experiment to address the performance and modeling-related challenges of these different datasets. An experimental evaluation of the different bankruptcy models is beyond the scope of this survey. We present in the followings a complementary, yet equally important assessment to address the quality and the informativeness of datasets as the function of their characteristics.

\subsection{Metrics for evaluating the quality of datasets}\label{sec3.1}

Although both machine learning and deep learning models are becoming very powerful for bankruptcy prediction, one needs to carefully consider which datasets are indeed applicable for the analysis. Not every datasets are suitable for training and provide the solution to the modeling objective. Gudivada et al. focus on data quality in the context of big data and machine learning, and list the dimensions for data quality assessment\cite{gudivada2017data}. As they argue, if the dataset is small, the model may overfit. If the paper operates with high dimensional dataset, the deep learning model may not be able to perform effective feature extraction. If there is no correlation in the data, the deep learning model may not be able to perform effective feature extraction on the data. If there are outliers in the dataset, the deep learning model may predict them incorrectly. 

These are commonly known analytical rules for all the datasets. However, the datasets used for bankruptcy prediction, as our taxonomy also conceptualizes, has their own characteristics. Real-world bankruptcy datasets often exhibit a significant imbalance between the number of bankrupt and non-bankrupt enterprises, with bankrupt enterprises typically being a minority. This imbalance poses challenges for model training and evaluation. Some studies have used datasets with artificially higher bankruptcy rates\cite{choe_garas_2021, Cultrera_2016, Liang_Lu_Tsai_Shih_2016}, which, although not reflective of real-world distributions, do not invalidate the conclusions of these studies. Understanding the quality of the dataset is essential for its proper use. Therefore, we propose a set of metrics to measure the quality of a dataset, ensuring that it is suitable for effective bankruptcy prediction (see Table\ref{quality}).

\begin{table}[h]
\caption{Metrics for dataset quality evaluation}\label{quality}
\begin{tabular}{@{}lp{3cm}p{6cm}@{}}
\toprule
\textbf{Evaluation Criteria} & \textbf{Metric} & \textbf{Description} \\
\midrule
Data Balance & Bankruptcy Rate & Check if the number of samples in the positive and negative classes are balanced. Imbalanced datasets can cause the model to be biased towards the class with more samples. \\
Data Volume & Sample Size, Number of Features & Assess whether the number of samples and features in the dataset is sufficient and balanced. Generally, more data can improve the model's generalization ability. \\
Data Integrity & Missing Values & Check if there are any missing values in the dataset. If present, determine if the proportion of missing values is acceptable and how to handle these missing values. \\
Data Noise & Noise Level & Evaluate the noise level in the data, i.e., whether the data contains incorrect labels or outliers. Noise can affect the model's training performance. \\
Data Distribution & Feature Distribution & Check the distribution of each feature to see if there are severe skews or outliers. \\
Data Redundancy & Duplicate Samples & Detect if there are duplicate samples in the dataset. Duplicate samples may lead to model overfitting. \\
\hline
\end{tabular}
\end{table}

\subsection{Metrics for evaluating the informativeness of datasets}\label{sec3.2}
In the financial industry, regulatory compliance necessitates high interpretability for  both features and models. This requirement means that each input feature in a model must have a clear and understandable meaning to explain the prediction outcomes effectively. Techniques like Principal Component Analysis (PCA) transform original features into new principal components, which are often difficult to interpret, thus reducing the model’s transparency and interpretability. Consequently, most bankruptcy prediction models do not employ dimensionality reduction methods, such as PCA and factor analysis, for feature engineering. However, the performance of models can sometimes degrade due to the presence of too many redundant features. Therefore, it is crucial to understand the informativeness of these features within their context. To address this need, we propose a set of metrics to evaluate the informativeness of features.  We also conduct experiments using available datasets to demonstrate the applicability and effectiveness of these metrics in assessing the quality of features for bankruptcy prediction.

Our experiment is inspired by existing literature on feature selection. Laborda et al.~\cite{laborda2021feature} mention using Chi-Squared test and correlation coefficients to select features for a credit scoring model. Trivedi et al.~\cite{trivedi2020study} conclude that the combination of random forest and Chi-squared tests is the best pair for building credit scoring models, over-performing tests such as information-gain or gain-ratio. Jemai et al.~\cite{jemai2023feature} mention the univariate feature selection method by calculating feature importance, information value, as well as Chi-squared tests can represent the correlation between features and labels. 
Ramya and Kumaresan~\cite{ramya2015analysis} point out that the effects of these, above-mentioned metrics are, in fact, similar. Based on these commonalities and findings, we deploy therefore for our experiment the information value, feature importance and Chi-squared tests to evaluate feature effects on informativeness. We shortly introduce the selected tests in the following subsections.

\subsubsection{Information Value}\label{subsubsec3.2.1}
Information Value (IV) is a statistical measure commonly used in credit scoring and predictive modeling to evaluate the predictive power of a variable. It quantifies the strength of the relationship between an independent variable and the dependent variable by measuring the variable's ability to differentiate between the categories of the dependent variable.

IV is calculated as follows:

\[ IV = \sum_{i=1}^{k} (P_{i0} - P_{i1}) \cdot \ln\left(\frac{P_{i0}}{P_{i1}}\right) \]

where \( k \) is the number of categories or bins of the independent variable, \( P_{i0} \) is the proportion of non-events (refers to active companies in this context) in the \( i \)th category, and \( P_{i1} \) is the proportion of events (refers to bankrupted companies in this context) in the \( i \)th category.

A higher IV indicates a stronger predictive power of the variable. Typically, if \( IV \geq 0.5 \), researchers associate a variable with a very strong predictive power. If \( 0.1 \leq IV < 0.3 \), a moderate predictive power is associated to the variable. Consequently, \( IV < 0.02 \) indicates no predictive power.
In practice, IV is often used during the variable selection process in predictive modeling to identify and retain the most influential variables for model development.

\subsubsection{Feature Importance}\label{subsubsec3.2.2}
Feature Importance (FI) is a crucial concept in machine learning and data analysis, aimed at identifying and quantifying the contribution of individual features or variables in a predictive model. Understanding FI helps practitioners interpret model outputs and prioritize variables based on their impact on the model's performance.

One commonly used method to calculate FI is the Mean Decrease in Impurity (MDI) for decision tree-based models, such as Random Forests. The formula for MDI can be expressed as follows:
\[ 
\begin{split}
    \text{MDI}(X_j) &= \sum_{t} p(t) \cdot \text{impurity}(t) \\
        &\quad - \sum_{t_L} p(t_L) \cdot \text{impurity}(t_L) \\
        &\quad - \sum_{t_R} p(t_R) \cdot \text{impurity}(t_R) 
\end{split}
\]

where \( X_j \) represents the j-th feature; \( t \) denotes a node in the tree; \( t_L \) and \( t_R \) represent the left and right child nodes, respectively; \( p(t) \), \( p(t_L) \), and \( p(t_R) \) are the proportions of samples reaching node \( t \), \( t_L \), and \( t_R \) among the total samples, respectively; and \( \text{impurity}() \) measures the impurity of a node, and it depends on the specific impurity criterion used (e.g., Gini impurity, entropy).

\( \text{FI}_{X_j} \) can be computed by normalizing the MDI values:

\[ \text{FI}_{X_j} = \frac{\text{MDI}(X_j)}{\sum_{j} \text{MDI}(X_j)} \]

The higher the FI value for a specific variable is, the more influential that variable is in the model. It is important to note that there are various other methods for calculating feature importance, depending on the model type and the underlying algorithms used.

\subsubsection{Chi-squared test}\label{subsubsec3.2.3}
The chi-squared (\( \chi^2 \)) test is a statistical method used to determine if there exists a significant association between categorical variables. It is particularly useful for analyzing contingency tables, which display the frequency distribution of two or more categorical variables.

The chi-squared test applies the following formula:

\[ \chi^2 = \sum_{i=1}^{r} \sum_{j=1}^{c} \frac{\left(O_{ij} - E_{ij}\right)^2}{E_{ij}} \]

where \( \chi^2 \) is the chi-square test statistic, \( O_{ij} \) represents the observed frequency in the \( i \)-th row and \( j \)-th column of the contingency table, \( E_{ij} \) is the expected frequency in the \( i \)-th row and \( j \)-th column, calculated under the assumption of independence between the variables.

The chi-squared test helps assess whether the observed frequencies significantly deviate from the expected frequencies, allowing researchers to determine if there is a statistically significant association between the variables. Commonly used, the null hypothesis (\( H_0 \)) assumes independence between the variables, while the alternative hypothesis (\( H_1 \)) suggests a significant association. If the calculated chi-square test statistic is greater than the critical value from the chi-squared contingency table, or if the p-value is less than the chosen significance level (commonly 0.05), then the null hypothesis is rejected, indicating a significant association between the categorical variables.

\subsection{Experimental setup and results}\label{subsec3.3}
In order to get realistic results for our metrics evaluation, we divide the analyzed datasets into two separate sub-sets, into a training and a testing set, respectively. The training sets provide the fundamentals to assess the metrics value to describe the quality of data in the context of bankruptcy prediction . We set the ratio of training and testing sets to 70\%:30\% in our experiment, and applied for each dataset accordingly. During our experiment, we also randomly split each dataset five times, thus repeated the setting of ratios  and used eventually the aggregated 
average of those five repeats as the metrics value.

The summary of quality metrics for the evaluated datasets is presented in Table \ref{datasets}. It is noted that the Russian dataset exhibits a significantly higher bankruptcy rate compared to others, which may offer modeling advantages but could potentially deviate from real-world scenarios. Moreover, the sample size of the Russian dataset is notably small, presenting challenges for model training. Additionally, the Russian dataset demonstrates a relatively high feature-to-sample size ratio. In contrast, the American dataset as well as the Taiwanese dataset shows an opposite trend with a large sample size but fewer features. This imbalance between sample size and number of features could also pose a challenge when modeling. The five Polish datasets seem more balanced with a reasonable bankruptcy rate. Therefore, it's important for researchers and professional to know the quality of the dataset before using it.

\begin{table*}[h]
  \centering
  \caption{Summary of quality metrics for datasets}
  \begin{tabular}{@{}p{2.8cm}p{1.6cm}p{1.7cm}p{1.8cm}p{1.5cm}p{1.2cm}@{}}
    \toprule
    Datasets & \# of bankrupted & \# of non bankrupted & Bankruptcy rate & \# of features & \# of sample \\
    \midrule
     American dataset & 5220 & 73462 & 6.63\% & 18 & 78682 \\
     Russian dataset & 456 &2001 & 18.56\% &66 &2457 \\
     Taiwanese dataset & 220& 6599& 3.23\%  & 95& 6819 \\
     Polish\_1year & 271&6756&3.86\% &64&7027\\
     Polish\_2year & 400&9773&3.93\% &64&10173\\
     Polish\_3year & 495&10008&4.71\% &64&10503\\
     Polish\_4year & 515&9277&5.26\% &64&9792\\
     Polish\_5year & 410&5500&6.94\% &64&5910\\
    \hline
  \end{tabular}
  \label{datasets}
\end{table*}

\begin{figure}
  \centering
  \includegraphics[width=\linewidth]{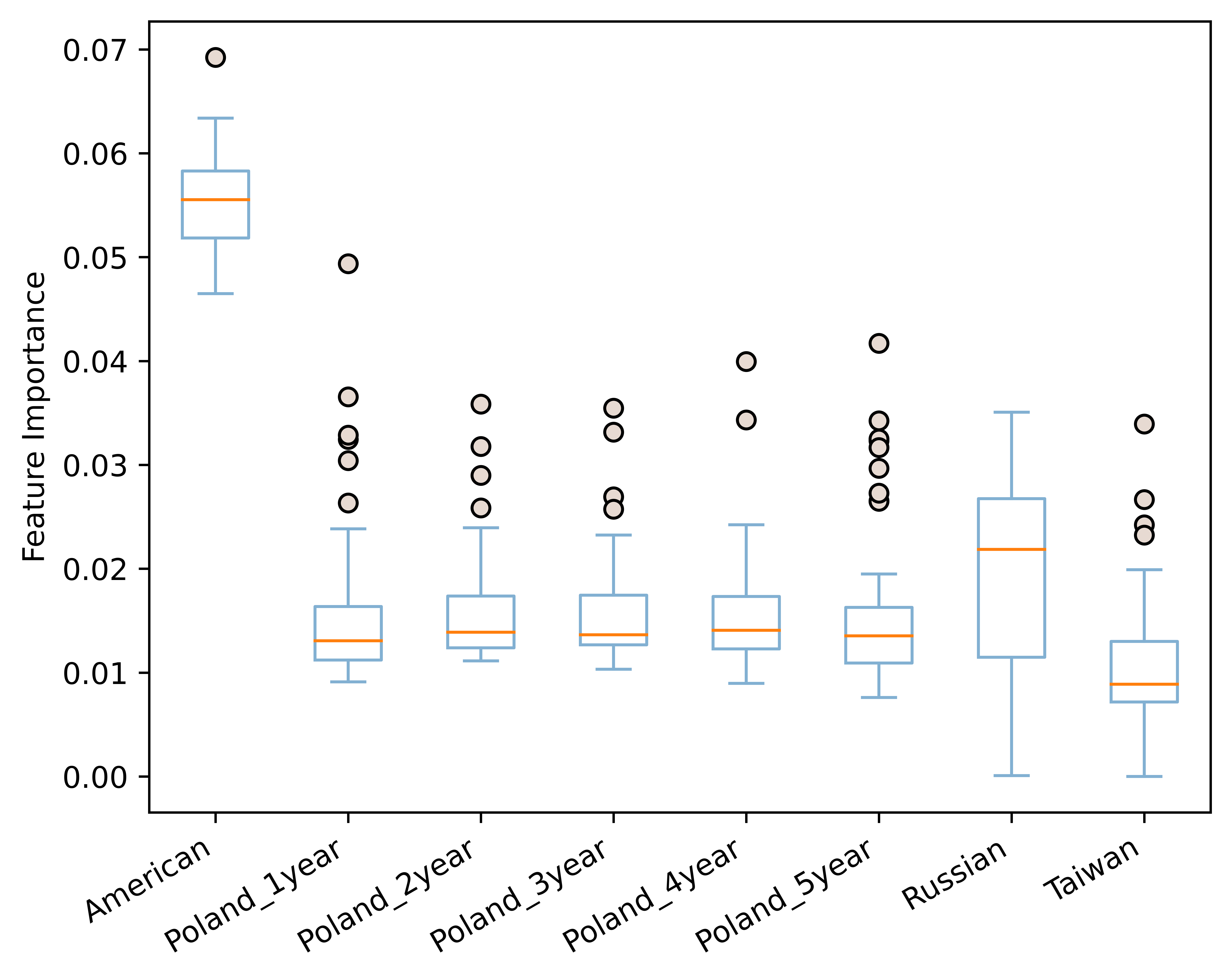}
  \caption{Feature importance of different datasets}
  \label{fi}
\end{figure}

\begin{figure}
  \centering
  \includegraphics[width=\linewidth]{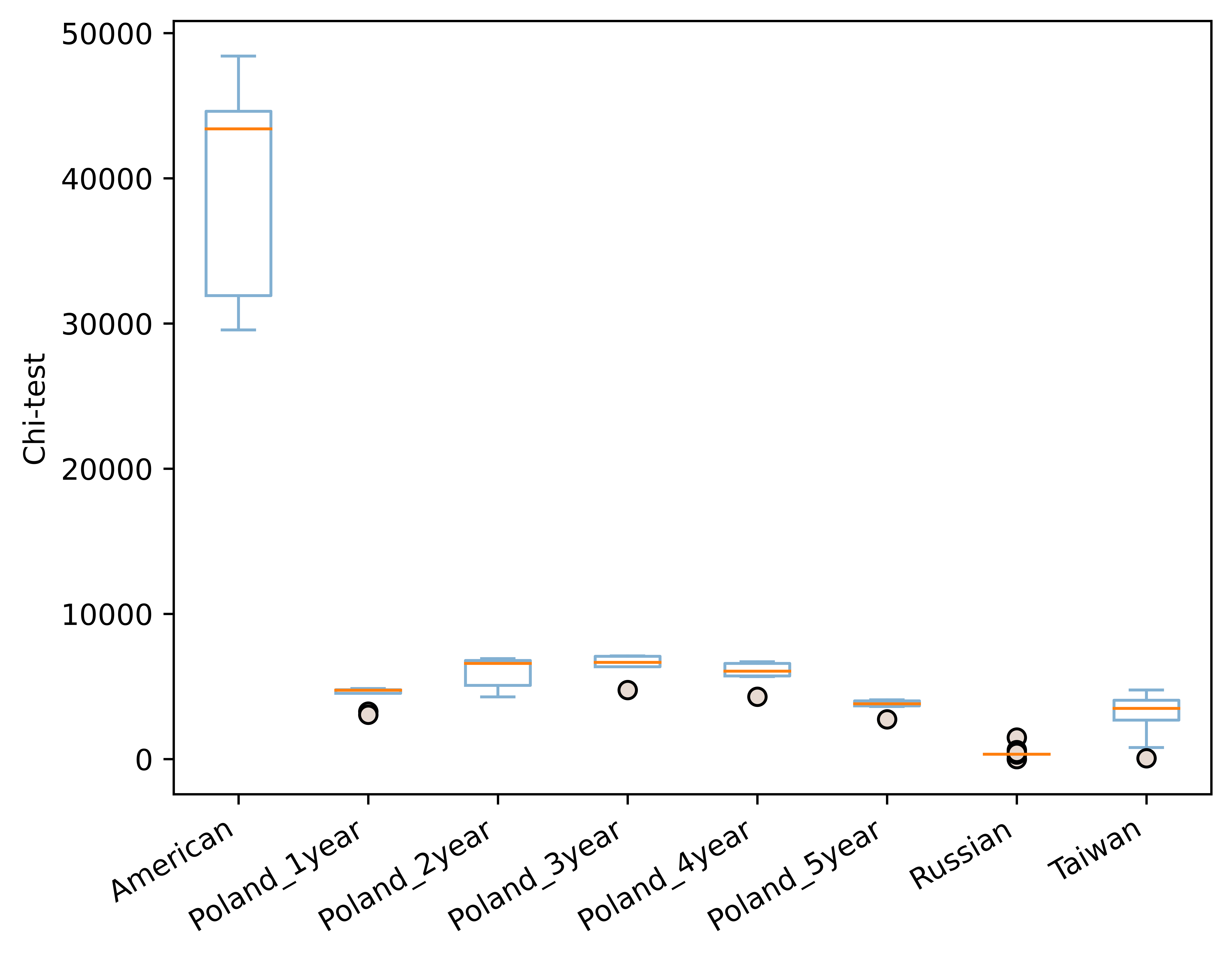}
  \caption{Chi-test value of different datasets}
  \label{chi}
\end{figure}

\begin{figure}
  \centering
  \includegraphics[width=\linewidth]{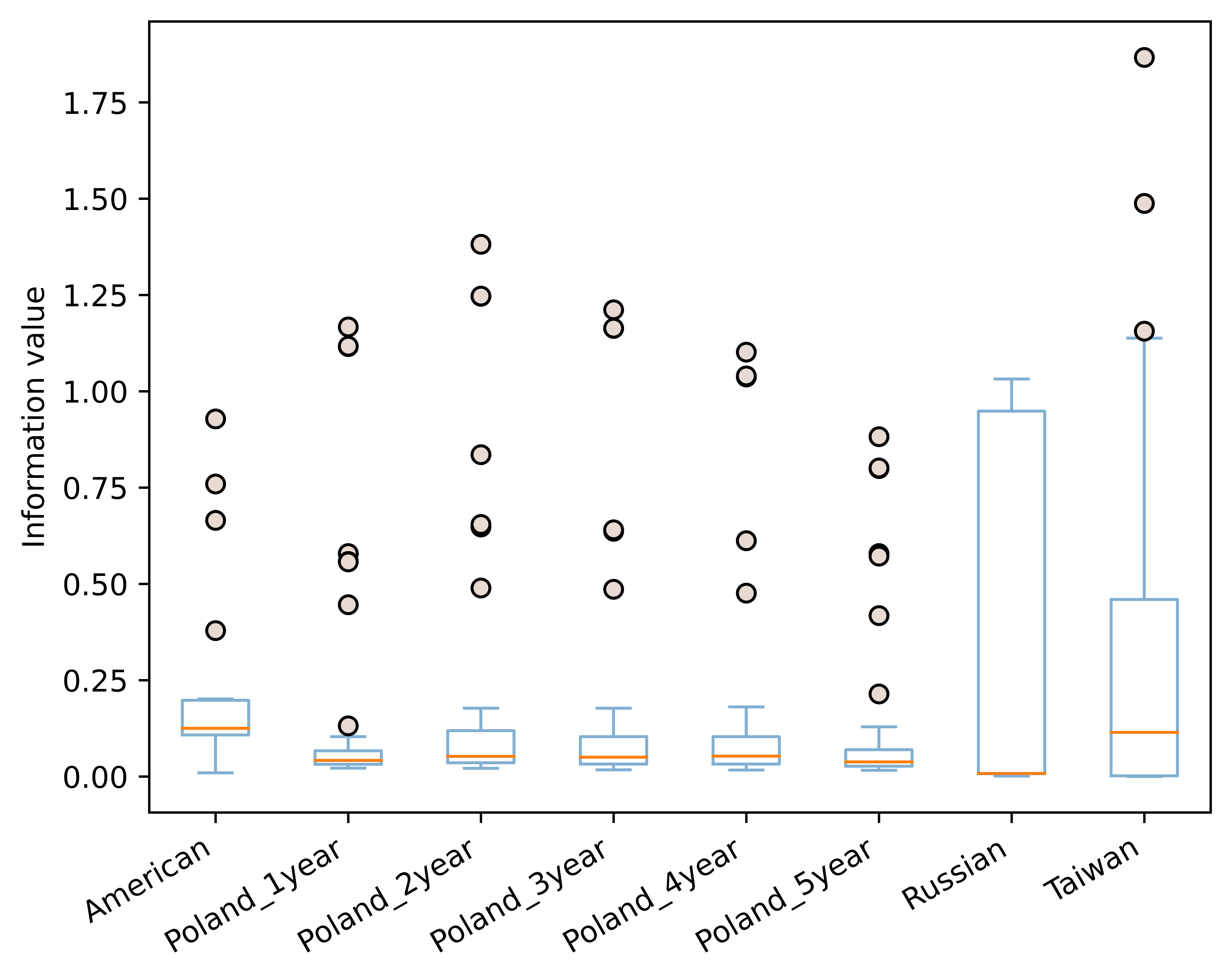}
  \caption{Information value of different datasets}
  \label{iv}
\end{figure}

Figures~\ref{fi}, ~\ref{chi} and ~\ref{iv} summarize our findings. We compare the metrics value of different datasets using box plots. As Figures~\ref{fi} and ~\ref{iv} indicate, the features of the American dataset perform the best for both feature importance and the Chi-squared test. However, it ranks relatively low for information value. The apparent difference of metrics rankings, as our experiment concludes, shows an opposite finding compared to what Ramya and Kumaresan\cite{ramya2015analysis} conclude. 

Features from Russian dataset have a relatively high value for feature importance but not that good for Chi-squared test. In terms of information value, features from Russian dataset speared over a wide range, which means it has features with extreme high information value and at the same time there are features with extreme low information value.

The performance of five Polish datasets are quite similar in each metric may because they share the same features. The average lines of each box representing for each Polish dataset are basically at the same level for these three datasets. All of them are ranked in the middle.

\section{Conclusion}\label{sec4}
Bankruptcy prediction has important practical applications in the financial field, and the application of diverse data sources enhances more comprehensive and multi-perspective information for constructing prediction models. From traditional financial data to emerging social media and alternative data, as well as incorporating machine learning techniques, all contribute to the accuracy and usefulness of bankruptcy prediction. However, rational integration and mining of these data, along with continuous improvement of predictive models, remain crucial issues to be addressed in future research and practice.

In this study, we collected different datasets used in this paper from 47 literatures related to bankruptcy prediction, and conducted qualitative and quantitative analysis on them, hoping to bring insight to researchers. Our findings indicate that while different data sources provide more information for bankruptcy forecasting, integrating this data and mining valuable information is challenging. Data quality, data inconsistency, and cross-data source issues are challenges that need to be overcome.
When applying bankruptcy prediction models constructed from different data sources, model evaluation and improvement are critical. Backtesting and validation of models using historical data, adoption of appropriate performance metrics to measure the accuracy of predictions, and improvement and optimization of models based on evaluation results.

At present, accounting-based data remains the predominant dataset for training bankruptcy prediction models. This kind of dataset is relatively easy to obtain and the data quality is high. At the same time, we also find that the new data type of relational dataset also plays a very important role in bankruptcy prediction, but this kind of data usually involves some sensitive data such as transaction information, and the knowledge graph of enterprises and people, which is not easy to obtain.

Although many datasets are public datasets, most of them are paid for or self-integrated. There are not many datasets that are completely free and can be downloaded directly. We will continue to follow this information in the future.

\section{Acknowledgement}
This work was supported by National Research Fund Luxembourg (FNR) under Grant 15403349 and Yoba S.A..

\begin{appendices}

\section{Resources for surveyed datasets}\label{secA1}

\subsection{Free access}

1. Cyprus Stock Exchange,\url{https://www.cse.com.cy/en-GB}

2. SMEsD, \url{https://github.com/shaopengw/comrisk}

3. HAT, \url{https://github.com/hetergraphforbankruptcypredict/HAT}

4. Russia dataset, \url{https://github.com/yzelenkov/classification_with_feature_selection/}

5. Polish dataset, \url{https://archive.ics.uci.edu/dataset/365/polish+companies+bankruptcy+data}

6. American dataset, \url{https://github.com/sowide/bankruptcy_dataset}

7. Taiwanese dataset, \url{https://archive.ics.uci.edu/dataset/572/taiwanese+bankruptcy+prediction}

8. Management Discussion and Analysis section of 10-K, can be found by the keywords of a company's name and "10-k" in \url{https://www.sec.gov/} from the search engine. For example, \url{https://www.sec.gov/Archives/edgar/data/27904/000002790416000018/dal1231201510k.htm} is a 10-k file of Delta Airline

\subsection{Subscription fee applied}

1. Albertina database, \url{https://www.aip.cz/en/}

2. Standard and Poor's Capital IQ service, \url{https://www.spglobal.com/en/}

3. Bureau Van Dijk,\url{https://www.bvdinfo.com/en-gb/}

4. Johannesburg Stock Exchange, \url{https://www.cmegroup.com/market-data/third-party-data-johannesburg-stock-exchange.html}

5. Infotel, \url{http://infotel.es}

6. Center for Research in Security Prices (CRSP) dataset, \url{https://www.crsp.org/}

7. Compustat/Compustat North America/Compustat Global, \url{https://www.marketplace.spglobal.com/en/datasets/compustat-financials-(8)}

8. Bloomberg, \url{https://www.bloomberg.com/}

9. CRIF-Slovak Credit Bureau, \url{https://www.crif.sk/}

10. Diane database of Bureau Van Dijk, \url{https://www.bvdinfo.com/en-gb/our-products/data/national/diane}

11. Orbis database of Bureau van Dijk, \url{https://www.bvdinfo.com/en-gb/our-products/data/international/orbis}

12. Taiwan Economic Journal, \url{http://www.tej.com.tw/twsite/}

13. Datastream, \url{https://www.lseg.com/en/data-analytics/products/datastream-macroeconomic-analysis}

14. Thomson One Banker, \url{https://banker.ib.thomsonone.com/}

15. London Share Price Database, \url{https://www.london.edu/faculty-and-research/finance/london-share-price-database}

16. Belfirst of Bureau Van Dijk, \url{https://www.bvdinfo.com/en-us/our-products/data/national/bel-first}

17. Fame of Bureau Van Dijk, \url{https://www.bvdinfo.com/en-gb/our-products/data/national/fame}

18. KOSME, \url{https://www.kosme.com/en/}

19. SABI of Bureau van Dijk, \url{https://login.bvdinfo.com/R0/SabiNeo}

20. AMADEUS of Bureau Van Dijk, \url{https://www.bvdinfo.com/en-gb/our-products/data/international/amadeus}

21. Taiwan Stock Exchange Corporation, \url{https://www.twse.com.tw/en/}

22. Serasa Experian, \url{https://www.serasaexperian.com.br/datalab/}

23. CERVED, \url{https://www.cerved.com/en/}

24. Federal Deposit Insurance Corporation, \url{https://www.fdic.gov/bank/statistical/index.html}

25. Securities Exchange Commission, \url{https://www.sec.gov/}

\section{Summary for related datasets}

\begin{center}
\begin{landscape}

\label{summary}
\begin{longtable}{p{1cm}p{1cm}p{1.5cm}p{1.8cm}p{1.5cm}p{4cm}p{3cm}p{1.5cm}p{1.5cm}}
\caption{Summary for related studies}\\
\toprule
Studies& Year of Publication & \# of Sample & Bankruptcy rate & \# of Features& Data Type& Data Source& Publicly available?& Fee applied?\\
\midrule
\endfirsthead

\multicolumn{9}{c}{Table VI continued} \\
\toprule
Studies& Year of Publication & \# of Sample & Bankruptcy rate & \# of Features& Data Type& Data Source& Publicly available?& Fee applied?\\
\midrule
\endhead

\hline
\multicolumn{9}{r}{{continued on next page}} \\
\endfoot

\hline
\endlastfoot

\cite{obradovic_2018}&2018&126& / &24& Accounting-based & Serbia & No&  /\\
\cite{stefko_2020} &2020&343& 25.36\% &9& Accounting-based & CRIF-Slovak Credit Bureau& No& / \\
\cite{nemec_pavlik_2016}&2016&2061& / &6& Accounting-based & Albertina database & Yes & Yes \\
\cite{Jones_2017}&2017&1115& 6.66\%&82& Accounting-based, Macroeconomic indicators, Corporate govenance indicators,Basic informationrmation& Standard and Poor's Capital IQ service & Yes & Yes \\
\cite{Cultrera_2016} &2016&7152& 50.00\% &9& Accounting-based, Basic information& Bureau Van Dijk& Yes & Yes \\
\cite{Lukason_Andresson_2019} &2019&4515& 11.34\% &14& Accounting-based,Tax-based & Estonian & No& \\
\cite{nouri_soltani_2016} &2016&318& 22.96\% &12& Accounting-based, Market-based, Macroeconomic indicators & Cyprus Stock Exchange& Yes & No\\
\cite{Sabela_Brummer_Hall_Wolmarans_2018} &2018&100& 22.00\% &35& Accounting-based, Market-based, Macroeconomic indicators & Johannesburg Stock Exchange& Yes & Yes \\
\cite{Faris_Abukhurma_Almanaseer_Saadeh_Mora_Castillo_Aljarah_2020}&2020&2860& 2.17\%&33& Accounting-based, Basic information& Infotel& Yes & Yes \\
\cite{Kim_Cho_Ryu_2022}&2022&454752& 0.45\%&8& Market-based & Center for Research in Security Prices (CRSP) dataset& Yes & Yes \\
\cite{Reisz_Perlich_2005} &2005&33037& 2.42\%&20& Market-based & Compustat& Yes & Yes \\
\cite{choe_garas_2021}&2020&84& 50.00\% &5& Accounting-based & Bloomberg& Yes & Yes \\
\cite{Manuel_2023} &2023&186& 9.68\%&5& Accounting-based & CRIF-Slovak Credit Bureau& Yes & Yes \\
\cite{zhao_wei_guo_yang_chen_li_zhuang_liu_kou_2022} &2024&889& / & / & Basic information,Law suit related, Knowledge graph of SMEs& SMEsD& Yes & No\\
\cite{zheng_lee_wu_pan_2021}&2021&13489& 26.44\% & / & Network of board member and shareholder& HAT& Yes & No\\
\cite{stefko_2021} &2021&498& 19.48\% & / & Accounting-based & CRIF-Slovak Credit Bureau& Yes & Yes \\

\multirow{2}{*}{\cite{zelenkov2017two},\cite{Zelenkov_Volodarskiy_2021}} & \multirow{2}{=}{2017, 2021} &2457& 18.56\% &55& Accounting-based, Business environment factors & Russia dataset & Yes & No\\
 & &5910& 6.94\%&63& Accounting-based and business environment factors& Polish dataset & Yes & No\\

\multirow{3}{*}{\cite{kou_xu_peng_shen_chen_chang_kou_2021}} & \multirow{3}{*}{2021} &168466& 0.64\%& \multirow{3}{*}{98} & \multirow{3}{=}{Basic information, Knowledge graph of SMEs, Transaction-based, Payment network–based}& \multirow{3}{=}{Business Bank of Shandong} & \multirow{3}{=}{No} & \multirow{3}{=}{ /} \\
 & &167364& 0.48\%& &&& & \\
 & &166527& 0.26\%& &&& & \\
&&& &&&&&\\

\cite{Mendes_Cardoso_2014}&2014&2033& 22.92\% &39& Accounting-based & ANS& No& / \\
\cite{Le_Lee_Park_Baik_2018}&2018&120355& 0.26\%&38& Accounting-based, Transactional-based variables& Korean financial company & No& / \\
\cite{Chen_Ribeiro_Vieira_Duarte_Neves_2011} &2011&1200& 50.00\% &30& Accounting-based & Diane database of Bureau Van Dijk& Yes& Yes\\
\cite{doumpos_andriosopoulos_galariotis_makridou_zopounidis_2017}&2017&138387& 1.29\%&30& Basic information,Macroeconomic indicators, energy & Orbis database of Bureau van Dijk& Yes & Yes \\
\cite{Lombardo_Pellegrino_Adosoglou_Cagnoni_Pardalos_Poggi_2022}&2022&78682& 0.77\%&18& Accounting-based & American dataset& Yes & No\\
\cite{Mai_Tian_Lee_Ma_2019} &2019&94994& 0.50\%&36& Accounting-based, Market-based, Text from annual reports & Compustat North America, Securities Exchange Commission, Management Discussion and Analysis section of 10-K & Yes & Yes \\
\cite{Liang_Lu_Tsai_Shih_2016}&2016&478& 50.00\% &190& Accounting-based, Company goverance indicators& Taiwan Economic Journal& Yes & Yes \\
\cite{Li_Miu_2009}&2009&6288& 21.14\% & & Accounting-based, Market-based & Compustat& Yes & Yes \\
\cite{hernandez_tinoco_wilson_2013}&2013&23218& 5.40\%&10& Accounting-based, Market-based, Macroeconomic indicators & Datastream,Thomson One Banker and London Share Price Database& Yes & Yes \\
\cite{tobback2017bankruptcy} &2017&2400000& / &6& Accounting-based, Relational data& Belfirst and Fame databases of Bureau Van Dijk & Yes & Yes \\
\multirow{2}{*}{\cite{Zhou_2013}}& \multirow{2}{*}{2013} &86129& 1.07\%&10& Accounting-based & Compustat North America & Yes & Yes \\
 & &36637& 0.16\%&10& accounting & Compustat Global& Yes & Yes \\
\cite{du_jardin_2015} &2015& /& / &50& Accounting-based & Diane database of Bureau Van Dijk& Yes & Yes \\
\cite{Lee_Choi_Yoo_2020} &2020&4358& 26.78\% &37& Management ability, Business feasibility, Technical ability, other & KOSME& Yes & Yes \\
\cite{Pervan_Kuvek_2013}&2013&825& 15.39\% &22& Financial data, Non-financial data(Firm owners' personal credit peRandom forestormance, management quality etc.) & Croatian commercial bank & No& / \\
\cite{lasheras_2012} &2012&63107& 0.41\%&5& Accounting-based & SABI of Bureau van Dijk & Yes & Yes \\

\multirow{2}{*}{\cite{Svabova_Michalkova_Durica_Nica_2020}} & \multirow{2}{*}{2020} &75652& 10.87\% & \multirow{2}{*}{11} & Accounting-based & \multirow{2}{=}{AMADEUS of Bureau Van Dijk} & \multirow{2}{*}{Yes}& \multirow{2}{*}{Yes}\\
 & &75652& 12.55\% & &&& & \\

\cite{Chuang_2013} &2013&321& 13.08\% &26& Accounting-based & Taiwan Stock Exchange Corporation and Taiwan Economic Journal& Yes & Yes \\
\cite{Fernandes_Artes_2016} &2016&9000000& / &19& Accounting-based, Spatial information& Serasa Experian& No& / \\
\cite{Figini_Bonelli_Giovannini_2017}&2017&38036& 5.00\%& / & Accounting-based, Basic information, Transactional-based & UniCredit bank & No& No\\

\multirow{2}{*}{\cite{Wang_Ma_Yang_2014}}& \multirow{2}{*}{2014} &240& 46.67\% &30& Accounting-based & Polish dataset & Yes & No\\
 & &132& 50.00\% &24& Accounting-based& CD-ROM of Data Mining for Business Intelligence: Concepts, Techniques, and Applications in Microsoft Office Excel with XLMiner & Yes & Yes \\

\cite{Ptak-Chmielewska_2019} &2019&806& 38.59\% & / & Accounting-based, Basic information& Polish consultancy firm& No& /\\
\multirow{2}{*}{\cite{Ciampi_2015}}& \multirow{2}{*}{2015} &3210& 50.00\% &38& Accounting-based & \multirow{2}{*}{CERVED}& Yes & Yes \\
 & &1156& 50.00\% &38& Accounting-based, Company goverance indicators && Yes & Yes \\

\cite{Kim_Byeong_Mun_Bae_Byeong_Mun_2018} &2018&288& 50.00\% &53& Accounting-based & KOSPI& No & / \\

\multirow{5}{*}{\cite{Alam_Shaukat_Mushtaq_Ali_Khushi_Luo_Wahab_2021}} & \multirow{5}{*}{2021} &7027& 3.86\%& \multirow{5}{*}{64} & \multirow{5}{*}{Accounting-based}& \multirow{5}{*}{Polish dataset}& \multirow{5}{*}{Yes}& \multirow{5}{*}{No} \\
 & &10173& 3.93\%& &&& & \\
 & &10503& 4.71\%& &&& & \\
 & &9792& 5.26\%& &&& & \\
 & &5910& 6.94\%& &&& & \\
\cite{Casado_2013} &2013&1000& 50.00\% &6& Accounting-based & AMADEUS of Bureau Van Dijk& Yes & Yes \\
\cite{Jo_Shin_2016} &2016&313& 33.23\% &232& Accounting-based,Sentiment lexicon/variables & KOSPI and KOSDAQ & No & / \\
\cite{Petropoulos_Siakoulis_Stavroulakis_Vlachogiannakis_2020}&2020&101641& 1.55\%&660& Accounting-based & Federal Deposit Insurance Corporation & Yes & No \\
\cite{son2019data}&2019&997940&2.32\%&13& Accounting-based, Basic information& NICE Information Service Co& No & / \\
\cite{altman2008value}&2008&5816021&1.15\%&43& Accounting-based,Basic information, Reported and compliance, Operational risk& /& No & / \\
\hline
  \end{longtable}
\end{landscape}

\end{center}

\end{appendices}

\bibliographystyle{elsarticle-num} 
\bibliography{reference}

\begin{thebibliography}{10}
\expandafter\ifx\csname url\endcsname\relax
  \def\url#1{\texttt{#1}}\fi
\expandafter\ifx\csname urlprefix\endcsname\relax\def\urlprefix{URL }\fi
\expandafter\ifx\csname href\endcsname\relax
  \def\href#1#2{#2} \def\path#1{#1}\fi

\bibitem{altman1994corporate}
E.~I. Altman, G.~Marco, F.~Varetto, Corporate distress diagnosis: Comparisons using linear discriminant analysis and neural networks (the italian experience), Journal of banking \& finance 18~(3) (1994) 505--529.

\bibitem{wu2010comparison}
Y.~Wu, C.~Gaunt, S.~Gray, A comparison of alternative bankruptcy prediction models, Journal of Contemporary Accounting \& Economics 6~(1) (2010) 34--45.

\bibitem{tsai2014comparative}
C.-F. Tsai, Y.-F. Hsu, D.~C. Yen, A comparative study of classifier ensembles for bankruptcy prediction, Applied Soft Computing 24 (2014) 977--984.

\bibitem{mihalovic2016performance}
M.~Mihalovic, Performance comparison of multiple discriminant analysis and logit models in bankruptcy prediction, Economics \& Sociology 9~(4) (2016) 101.

\bibitem{altman1968financial}
E.~I. Altman, Financial ratios, discriminant analysis and the prediction of corporate bankruptcy, The journal of finance 23~(4) (1968) 589--609.

\bibitem{lasheras_2012}
F.~Sánchez-Lasheras, J.~de~Andrés, P.~Lorca, F.~J. de~Cos~Juez, \href{http://dx.doi.org/10.1016/j.eswa.2012.01.135}{A hybrid device for the solution of sampling bias problems in the forecasting of firms’ bankruptcy}, Expert Systems with Applications (2012) 7512–7523\href {https://doi.org/10.1016/j.eswa.2012.01.135} {\path{doi:10.1016/j.eswa.2012.01.135}}.
\newline\urlprefix\url{http://dx.doi.org/10.1016/j.eswa.2012.01.135}

\bibitem{Svabova_Michalkova_Durica_Nica_2020}
L.~Svabova, L.~Michalkova, M.~Durica, E.~Nica, \href{http://dx.doi.org/10.3390/su12114572}{Business failure prediction for slovak small and medium-sized companies}, Sustainability (2020) 4572\href {https://doi.org/10.3390/su12114572} {\path{doi:10.3390/su12114572}}.
\newline\urlprefix\url{http://dx.doi.org/10.3390/su12114572}

\bibitem{Wang_Ma_Yang_2014}
G.~Wang, J.~Ma, S.~Yang, An improved boosting based on feature selection for corporate bankruptcy prediction (2014).

\bibitem{Chuang_2013}
C.-L. Chuang, \href{http://dx.doi.org/10.1016/j.ins.2013.02.015}{Application of hybrid case-based reasoning for enhanced performance in bankruptcy prediction}, Information Sciences (2013) 174–185\href {https://doi.org/10.1016/j.ins.2013.02.015} {\path{doi:10.1016/j.ins.2013.02.015}}.
\newline\urlprefix\url{http://dx.doi.org/10.1016/j.ins.2013.02.015}

\bibitem{Kim_Byeong_Mun_Bae_Byeong_Mun_2018}
S.~Kim, B.~M. Mun, S.~J. Bae, Data depth based support vector machines for predicting corporate bankruptcy, Applied Intelligence 48 (2018) 791--804.

\bibitem{Alam_Shaukat_Mushtaq_Ali_Khushi_Luo_Wahab_2021}
T.~M. Alam, K.~Shaukat, M.~Mushtaq, Y.~Ali, M.~Khushi, S.~Luo, A.~Wahab, \href{http://dx.doi.org/10.1093/comjnl/bxaa056}{Corporate bankruptcy prediction: An approach towards better corporate world}, The Computer Journal (2021) 1731–1746\href {https://doi.org/10.1093/comjnl/bxaa056} {\path{doi:10.1093/comjnl/bxaa056}}.
\newline\urlprefix\url{http://dx.doi.org/10.1093/comjnl/bxaa056}

\bibitem{Casado_2013}
A.~Callejón, A.~Casado, M.~Fernández, J.~Peláez, \href{http://dx.doi.org/10.1080/18756891.2013.754167}{A system of insolvency prediction for industrial companies using a financial alternative model with neural networks}, International Journal of Computational Intelligence Systems (2013) 29\href {https://doi.org/10.1080/18756891.2013.754167} {\path{doi:10.1080/18756891.2013.754167}}.
\newline\urlprefix\url{http://dx.doi.org/10.1080/18756891.2013.754167}

\bibitem{kou_xu_peng_shen_chen_chang_kou_2021}
G.~Kou, Y.~Xu, Y.~Peng, F.~Shen, Y.~Chen, K.~Chang, S.~Kou, Bankruptcy prediction for smes using transactional data and two-stage multiobjective feature selection, Decision Support Systems 140 (2021) 113429.
\newblock \href {https://doi.org/10.1016/j.dss.2020.113429} {\path{doi:10.1016/j.dss.2020.113429}}.

\bibitem{Jo_Shin_2016}
N.-O. Jo, K.-S. Shin, Bankruptcy prediction modeling using qualitative information based on big data analytics (2016).

\bibitem{zheng_lee_wu_pan_2021}
Y.~Zheng, V.~C. Lee, Z.~Wu, S.~Pan, Heterogeneous graph attention network for small and medium-sized enterprises bankruptcy prediction, in: Pacific-Asia Conference on Knowledge Discovery and Data Mining, Springer, 2021, pp. 140--151.

\bibitem{Lukason_Andresson_2019}
O.~Lukason, A.~Andresson, \href{http://dx.doi.org/10.3390/jrfm12040187}{Tax arrears versus financial ratios in bankruptcy prediction}, Journal of Risk and Financial Management (2019) 187\href {https://doi.org/10.3390/jrfm12040187} {\path{doi:10.3390/jrfm12040187}}.
\newline\urlprefix\url{http://dx.doi.org/10.3390/jrfm12040187}

\bibitem{shi2019overview}
Y.~Shi, X.~Li, \href{http://hdl.handle.net/2117/176066}{An overview of bankruptcy prediction models for corporate firms: A systematic literature review}, Intangible Capital 15~(2) (2019) 114--127.
\newblock \href {https://doi.org/https://doi.org/10.3926/ic.1354} {\path{doi:https://doi.org/10.3926/ic.1354}}.
\newline\urlprefix\url{http://hdl.handle.net/2117/176066}

\bibitem{kumar2007bankruptcy}
P.~R. Kumar, V.~Ravi, Bankruptcy prediction in banks and firms via statistical and intelligent techniques--a review, European journal of operational research 180~(1) (2007) 1--28.

\bibitem{kovacova2019systematic}
M.~Kovacova, T.~Kliestik, K.~Valaskova, P.~Durana, Z.~Juhaszova, Systematic review of variables applied in bankruptcy prediction models of visegrad group countries, Oeconomia Copernicana 10~(4) (2019) 743--772.

\bibitem{jones2023literature}
S.~Jones, A literature survey of corporate failure prediction models, Journal of Accounting Literature 45~(2) (2023) 364--405.

\bibitem{clement2020machine}
C.~Clement, et~al., Machine learning in bankruptcy prediction--a review, Journal of Public Administration, Finance and Law~(17) (2020) 178--196.

\bibitem{appiah2015predicting}
K.~O. Appiah, A.~Chizema, J.~Arthur, Predicting corporate failure: a systematic literature review of methodological issues, International Journal of Law and Management 57~(5) (2015) 461--485.

\bibitem{alaka2018systematic}
H.~A. Alaka, L.~O. Oyedele, H.~A. Owolabi, V.~Kumar, S.~O. Ajayi, O.~O. Akinade, M.~Bilal, Systematic review of bankruptcy prediction models: Towards a framework for tool selection, Expert Systems with Applications 94 (2018) 164--184.

\bibitem{bellovary2007review}
J.~L. Bellovary, D.~E. Giacomino, M.~D. Akers, \href{http://www.jstor.org/stable/41948574}{A review of bankruptcy prediction studies: 1930 to present}, Journal of Financial Education 33 (2007) 1--42.
\newline\urlprefix\url{http://www.jstor.org/stable/41948574}

\bibitem{chouhan2014predicting}
V.~Chouhan, B.~Chandra, S.~Goswami, Predicting financial stability of select bse companies revisiting altman z score, International Letters of Social and Humanistic Sciences 15~(2) (2014) 92--105.

\bibitem{anjum2012business}
S.~Anjum, Business bankruptcy prediction models: A significant study of the altman’s z-score model, Available at SSRN 2128475 (2012).

\bibitem{hauschild2013altman}
D.~Hauschild, Altman z-score: Not just for bankruptcy, From Z-score to" Green Zone" survivability: AMPros Corporation (2013).

\bibitem{cindik2021revision}
Z.~C{\i}nd{\i}k, I.~H. Armutlulu, A revision of altman z-score model and a comparative analysis of turkish companies’ financial distress prediction, National Accounting Review 3~(2) (2021) 237--255.

\bibitem{platt2006understanding}
H.~D. Platt, M.~B. Platt, Understanding differences between financial distress and bankruptcy, Review of Applied Economics 2~(1076-2016-87135) (2006) 141--157.

\bibitem{puro2019financial}
N.~Puro, D.~Nancy~Borkowski, L.~Hearld, N.~Carroll, J.~Byrd, D.~Smith, A.~Ghiasi, et~al., Financial distress and bankruptcy prediction: a comparison of three financial distress prediction models in acute care hospitals, Journal of Health Care Finance (2019).

\bibitem{lombardo2022machine}
G.~Lombardo, M.~Pellegrino, G.~Adosoglou, S.~Cagnoni, P.~M. Pardalos, A.~Poggi, Machine learning for bankruptcy prediction in the american stock market: Dataset and benchmarks, Future Internet 14~(8) (2022) 244.

\bibitem{misc_polish_companies_bankruptcy_data_365}
S.~Tomczak, {Polish companies bankruptcy data}, UCI Machine Learning Repository, {DOI}: https://doi.org/10.24432/C5F600 (2016).

\bibitem{misc_taiwanese_bankruptcy_prediction_572}
{Taiwanese Bankruptcy Prediction}, UCI Machine Learning Repository (2020).

\bibitem{zhao_wei_guo_yang_chen_li_zhuang_liu_kou_2022}
Y.~Zhao, S.~Wei, Y.~Guo, Q.~Yang, X.~Chen, Q.~Li, F.~Zhuang, J.~Liu, G.~Kou, Combining intra-risk and contagion risk for enterprise bankruptcy prediction using graph neural networks (Feb 2022).

\bibitem{tobback2017bankruptcy}
E.~Tobback, T.~Bellotti, J.~Moeyersoms, M.~Stankova, D.~Martens, Bankruptcy prediction for smes using relational data, Decision Support Systems 102 (2017) 69--81.

\bibitem{kou2021bankruptcy}
G.~Kou, Y.~Xu, Y.~Peng, F.~Shen, Y.~Chen, K.~Chang, S.~Kou, Bankruptcy prediction for smes using transactional data and two-stage multiobjective feature selection, Decision Support Systems 140 (2021) 113429.

\bibitem{le2018oversampling}
T.~Le, M.~Y. Lee, J.~R. Park, S.~W. Baik, Oversampling techniques for bankruptcy prediction: Novel features from a transaction dataset, Symmetry 10~(4) (2018) 79.

\bibitem{altman2015financial}
E.~I. Altman, M.~Iwanicz-Drozdowska, E.~K. Laitinen, A.~Suvas, Financial and non-financial variables as long-horizon predictors of bankruptcy, Available at SSRN 2669668 (2015).

\bibitem{mai2019deep}
F.~Mai, S.~Tian, C.~Lee, L.~Ma, Deep learning models for bankruptcy prediction using textual disclosures, European journal of operational research 274~(2) (2019) 743--758.

\bibitem{Ciampi_2015}
F.~Ciampi, \href{http://dx.doi.org/10.1016/j.jbusres.2014.10.003}{Corporate governance characteristics and default prediction modeling for small enterprises. an empirical analysis of italian firms}, Journal of Business Research (2015) 1012–1025\href {https://doi.org/10.1016/j.jbusres.2014.10.003} {\path{doi:10.1016/j.jbusres.2014.10.003}}.
\newline\urlprefix\url{http://dx.doi.org/10.1016/j.jbusres.2014.10.003}

\bibitem{liang2016financial}
D.~Liang, C.-C. Lu, C.-F. Tsai, G.-A. Shih, Financial ratios and corporate governance indicators in bankruptcy prediction: A comprehensive study, European journal of operational research 252~(2) (2016) 561--572.

\bibitem{gudivada2017data}
V.~Gudivada, A.~Apon, J.~Ding, Data quality considerations for big data and machine learning: Going beyond data cleaning and transformations, International Journal on Advances in Software 10~(1) (2017) 1--20.

\bibitem{choe_garas_2021}
K.~Choe, S.~Garas, The graph theoretical approach to bankruptcy prediction, The Journal of Accounting and Management,The Journal of Accounting and Management (Jan 2021).

\bibitem{Cultrera_2016}
L.~Cultrera, X.~Brédart, \href{http://dx.doi.org/10.1108/raf-06-2014-0059}{Bankruptcy prediction: the case of belgian smes}, Review of Accounting and Finance (2016) 101–119\href {https://doi.org/10.1108/raf-06-2014-0059} {\path{doi:10.1108/raf-06-2014-0059}}.
\newline\urlprefix\url{http://dx.doi.org/10.1108/raf-06-2014-0059}

\bibitem{Liang_Lu_Tsai_Shih_2016}
D.~Liang, C.-C. Lu, C.-F. Tsai, G.-A. Shih, Financial ratios and corporate governance indicators in bankruptcy prediction: A comprehensive study, European Journal of Operational Research 252~(2) (2016) 561–572.
\newblock \href {https://doi.org/10.1016/j.ejor.2016.01.012} {\path{doi:10.1016/j.ejor.2016.01.012}}.

\bibitem{laborda2021feature}
J.~Laborda, S.~Ryoo, Feature selection in a credit scoring model, Mathematics 9~(7) (2021) 746.

\bibitem{trivedi2020study}
S.~K. Trivedi, A study on credit scoring modeling with different feature selection and machine learning approaches, Technology in Society 63 (2020) 101413.

\bibitem{jemai2023feature}
J.~Jemai, A.~Zarrad, Feature selection engineering for credit risk assessment in retail banking, Information 14~(3) (2023) 200.

\bibitem{ramya2015analysis}
R.~Ramya, S.~Kumaresan, Analysis of feature selection techniques in credit risk assessment, in: 2015 International conference on advanced computing and communication systems, IEEE, 2015, pp. 1--6.

\bibitem{obradovic_2018}
D.~Bešlić~Obradović, D.~Jakšić, I.~Bešlić~Rupić, M.~Andrić, \href{http://dx.doi.org/10.1080/1331677x.2017.1421990}{Insolvency prediction model of the company: the case of the republic of serbia}, Economic Research-Ekonomska Istraživanja (2018) 139–157\href {https://doi.org/10.1080/1331677x.2017.1421990} {\path{doi:10.1080/1331677x.2017.1421990}}.
\newline\urlprefix\url{http://dx.doi.org/10.1080/1331677x.2017.1421990}

\bibitem{stefko_2020}
R.~Štefko, J.~Horváthová, M.~Mokrišová, \href{http://dx.doi.org/10.3390/jrfm13090212}{Bankruptcy prediction with the use of data envelopment analysis: An empirical study of slovak businesses}, Journal of Risk and Financial Management (2020) 212\href {https://doi.org/10.3390/jrfm13090212} {\path{doi:10.3390/jrfm13090212}}.
\newline\urlprefix\url{http://dx.doi.org/10.3390/jrfm13090212}

\bibitem{nemec_pavlik_2016}
D.~Němec, M.~Pavlík, Predicting insolvency risk of the czech companies (Jan 2016).

\bibitem{Jones_2017}
S.~Jones, \href{http://dx.doi.org/10.1007/s11142-017-9407-1}{Corporate bankruptcy prediction: a high dimensional analysis}, Review of Accounting Studies (2017) 1366–1422\href {https://doi.org/10.1007/s11142-017-9407-1} {\path{doi:10.1007/s11142-017-9407-1}}.
\newline\urlprefix\url{http://dx.doi.org/10.1007/s11142-017-9407-1}

\bibitem{nouri_soltani_2016}
B.~Nouri, M.~Soltani, Designing a bankruptcy prediction model based on account, market and macroeconomic variables (case study: Cyprus stock exchange), Iranian Journal of Management Studies,Iranian Journal of Management Studies (Jan 2016).

\bibitem{Sabela_Brummer_Hall_Wolmarans_2018}
S.~Sabela, L.~Brummer, J.~Hall, H.~Wolmarans, Using fundamental, market and macroeconomic variables to predict financial distress: A study of companies listed on the johannesburg stock exchange, Journal of Economic and Financial Sciences 11~(1) (Apr 2018).
\newblock \href {https://doi.org/10.4102/jef.v11i1.168} {\path{doi:10.4102/jef.v11i1.168}}.

\bibitem{Faris_Abukhurma_Almanaseer_Saadeh_Mora_Castillo_Aljarah_2020}
H.~Faris, R.~Abukhurma, W.~Almanaseer, M.~Saadeh, A.~M. Mora, P.~A. Castillo, I.~Aljarah, \href{http://dx.doi.org/10.1007/s13748-019-00197-9}{Improving financial bankruptcy prediction in a highly imbalanced class distribution using oversampling and ensemble learning: a case from the spanish market}, Progress in Artificial Intelligence (2020) 31–53\href {https://doi.org/10.1007/s13748-019-00197-9} {\path{doi:10.1007/s13748-019-00197-9}}.
\newline\urlprefix\url{http://dx.doi.org/10.1007/s13748-019-00197-9}

\bibitem{Kim_Cho_Ryu_2022}
H.~Kim, H.~Cho, D.~Ryu, \href{http://dx.doi.org/10.1007/s10614-021-10126-5}{Corporate bankruptcy prediction using machine learning methodologies with a focus on sequential data}, Computational Economics (2022) 1231–1249\href {https://doi.org/10.1007/s10614-021-10126-5} {\path{doi:10.1007/s10614-021-10126-5}}.
\newline\urlprefix\url{http://dx.doi.org/10.1007/s10614-021-10126-5}

\bibitem{Reisz_Perlich_2005}
A.~Reisz, C.~Perlich, \href{http://dx.doi.org/10.2139/ssrn.531342}{A market-based framework for bankruptcy prediction}, SSRN Electronic Journal (Jul 2005).
\newblock \href {https://doi.org/10.2139/ssrn.531342} {\path{doi:10.2139/ssrn.531342}}.
\newline\urlprefix\url{http://dx.doi.org/10.2139/ssrn.531342}

\bibitem{Manuel_2023}
J.~Manuel, J.~Horváthová, M.~Mokrišová, M.~Bača, Bankruptcy prediction for sustainability of businesses: The application of graph theoretical modeling (2023).

\bibitem{stefko_2021}
R.~Štefko, J.~Horváthová, M.~Mokrišová, \href{http://dx.doi.org/10.3390/jrfm14050220}{The application of graphic methods and the dea in predicting the risk of bankruptcy}, Journal of Risk and Financial Management (2021) 220\href {https://doi.org/10.3390/jrfm14050220} {\path{doi:10.3390/jrfm14050220}}.
\newline\urlprefix\url{http://dx.doi.org/10.3390/jrfm14050220}

\bibitem{zelenkov2017two}
Y.~Zelenkov, E.~Fedorova, D.~Chekrizov, Two-step classification method based on genetic algorithm for bankruptcy forecasting, Expert Systems with Applications 88 (2017) 393--401.

\bibitem{Zelenkov_Volodarskiy_2021}
Y.~Zelenkov, N.~Volodarskiy, Bankruptcy prediction on the base of the unbalanced data using multi-objective selection of classifiers, Expert Systems with Applications 185 (2021) 115559.
\newblock \href {https://doi.org/10.1016/j.eswa.2021.115559} {\path{doi:10.1016/j.eswa.2021.115559}}.

\bibitem{Mendes_Cardoso_2014}
A.~Mendes, R.~L. Cardoso, P.~C. Mário, A.~L. Martinez, F.~R. Ferreira, \href{http://dx.doi.org/10.1002/isaf.1352}{Insolvency prediction in the presence of data inconsistencies}, Intelligent Systems in Accounting, Finance and Management (2014) 155–167\href {https://doi.org/10.1002/isaf.1352} {\path{doi:10.1002/isaf.1352}}.
\newline\urlprefix\url{http://dx.doi.org/10.1002/isaf.1352}

\bibitem{Le_Lee_Park_Baik_2018}
T.~Le, M.~Lee, J.~Park, S.~Baik, Oversampling techniques for bankruptcy prediction: Novel features from a transaction dataset, Symmetry 10~(4) (2018) 79.
\newblock \href {https://doi.org/10.3390/sym10040079} {\path{doi:10.3390/sym10040079}}.

\bibitem{Chen_Ribeiro_Vieira_Duarte_Neves_2011}
N.~Chen, B.~Ribeiro, A.~S. Vieira, J.~Duarte, J.~C. Neves, \href{http://dx.doi.org/10.1016/j.eswa.2011.04.090}{A genetic algorithm-based approach to cost-sensitive bankruptcy prediction}, Expert Systems with Applications (2011) 12939–12945\href {https://doi.org/10.1016/j.eswa.2011.04.090} {\path{doi:10.1016/j.eswa.2011.04.090}}.
\newline\urlprefix\url{http://dx.doi.org/10.1016/j.eswa.2011.04.090}

\bibitem{doumpos_andriosopoulos_galariotis_makridou_zopounidis_2017}
M.~Doumpos, K.~Andriosopoulos, E.~Galariotis, G.~Makridou, C.~Zopounidis, Corporate failure prediction in the european energy sector: A multicriteria approach and the effect of country characteristics, Research Papers in Economics,Research Papers in Economics (Jan 2017).

\bibitem{Lombardo_Pellegrino_Adosoglou_Cagnoni_Pardalos_Poggi_2022}
G.~Lombardo, M.~Pellegrino, G.~Adosoglou, S.~Cagnoni, P.~M. Pardalos, A.~Poggi, \href{http://dx.doi.org/10.3390/fi14080244}{Machine learning for bankruptcy prediction in the american stock market: Dataset and benchmarks}, Future Internet (2022) 244\href {https://doi.org/10.3390/fi14080244} {\path{doi:10.3390/fi14080244}}.
\newline\urlprefix\url{http://dx.doi.org/10.3390/fi14080244}

\bibitem{Mai_Tian_Lee_Ma_2019}
F.~Mai, S.~Tian, C.~Lee, L.~Ma, Deep learning models for bankruptcy prediction using textual disclosures, European Journal of Operational Research 274~(2) (2019) 743–758.
\newblock \href {https://doi.org/10.1016/j.ejor.2018.10.024} {\path{doi:10.1016/j.ejor.2018.10.024}}.

\bibitem{Li_Miu_2009}
M.~Li, P.~Miu, A hybrid bankruptcy prediction model with dynamic loadings on accounting-ratio-based and market-based information: A binary quantile regression approach, Journal of Empirical Finance,Journal of Empirical Finance (Apr 2009).

\bibitem{hernandez_tinoco_wilson_2013}
M.~Hernandez~Tinoco, N.~Wilson, \href{http://dx.doi.org/10.1016/j.irfa.2013.02.013}{Financial distress and bankruptcy prediction among listed companies using accounting, market and macroeconomic variables}, International Review of Financial Analysis (2013) 394–419\href {https://doi.org/10.1016/j.irfa.2013.02.013} {\path{doi:10.1016/j.irfa.2013.02.013}}.
\newline\urlprefix\url{http://dx.doi.org/10.1016/j.irfa.2013.02.013}

\bibitem{Zhou_2013}
L.~Zhou, \href{http://dx.doi.org/10.1016/j.knosys.2012.12.007}{Performance of corporate bankruptcy prediction models on imbalanced dataset: The effect of sampling methods}, Knowledge-Based Systems (2013) 16–25\href {https://doi.org/10.1016/j.knosys.2012.12.007} {\path{doi:10.1016/j.knosys.2012.12.007}}.
\newline\urlprefix\url{http://dx.doi.org/10.1016/j.knosys.2012.12.007}

\bibitem{du_jardin_2015}
P.~du~Jardin, \href{http://dx.doi.org/10.1016/j.ejor.2014.09.059}{Bankruptcy prediction using terminal failure processes}, European Journal of Operational Research (2015) 286–303\href {https://doi.org/10.1016/j.ejor.2014.09.059} {\path{doi:10.1016/j.ejor.2014.09.059}}.
\newline\urlprefix\url{http://dx.doi.org/10.1016/j.ejor.2014.09.059}

\bibitem{Lee_Choi_Yoo_2020}
S.~Lee, K.~Choi, D.~Yoo, \href{http://dx.doi.org/10.3390/su12239790}{Predicting the insolvency of smes using technological feasibility assessment information and data mining techniques}, Sustainability (2020) 9790\href {https://doi.org/10.3390/su12239790} {\path{doi:10.3390/su12239790}}.
\newline\urlprefix\url{http://dx.doi.org/10.3390/su12239790}

\bibitem{Pervan_Kuvek_2013}
I.~Pervan, T.~Kuvek, The relative importance of financial ratios and nonfinancial variables in predicting of insolvency, Croatian Operational Research Review,Croatian Operational Research Review (Feb 2013).

\bibitem{Fernandes_Artes_2016}
G.~B. Fernandes, R.~Artes, \href{http://dx.doi.org/10.1016/j.ejor.2015.07.013}{Spatial dependence in credit risk and its improvement in credit scoring}, European Journal of Operational Research (2016) 517–524\href {https://doi.org/10.1016/j.ejor.2015.07.013} {\path{doi:10.1016/j.ejor.2015.07.013}}.
\newline\urlprefix\url{http://dx.doi.org/10.1016/j.ejor.2015.07.013}

\bibitem{Figini_Bonelli_Giovannini_2017}
S.~Figini, F.~Bonelli, E.~Giovannini, \href{http://dx.doi.org/10.1016/j.dss.2017.08.001}{Solvency prediction for small and medium enterprises in banking}, Decision Support Systems (2017) 91–97\href {https://doi.org/10.1016/j.dss.2017.08.001} {\path{doi:10.1016/j.dss.2017.08.001}}.
\newline\urlprefix\url{http://dx.doi.org/10.1016/j.dss.2017.08.001}

\bibitem{Ptak-Chmielewska_2019}
A.~Ptak-Chmielewska, \href{http://dx.doi.org/10.3390/jrfm12010030}{Predicting micro-enterprise failures using data mining techniques}, Journal of Risk and Financial Management (2019) 30\href {https://doi.org/10.3390/jrfm12010030} {\path{doi:10.3390/jrfm12010030}}.
\newline\urlprefix\url{http://dx.doi.org/10.3390/jrfm12010030}

\bibitem{Petropoulos_Siakoulis_Stavroulakis_Vlachogiannakis_2020}
A.~Petropoulos, V.~Siakoulis, E.~Stavroulakis, N.~Vlachogiannakis, Predicting bank insolvencies using machine learning techniques, International Journal of Forecasting 36~(3) (2020) 1092–1113.
\newblock \href {https://doi.org/10.1016/j.ijforecast.2019.11.005} {\path{doi:10.1016/j.ijforecast.2019.11.005}}.

\bibitem{son2019data}
H.~Son, C.~Hyun, D.~Phan, H.~J. Hwang, Data analytic approach for bankruptcy prediction, Expert Systems with Applications 138 (2019) 112816.

\bibitem{altman2008value}
E.~I. Altman, G.~Sabato, N.~Wilson, The value of non-financial information in sme risk management, Available at SSRN 1320612 (2008).

\end{thebibliography}

\end{document}